\documentclass[12pt,draftclsnofoot,onecolumn]{IEEEtran}
\usepackage{amssymb}
\usepackage{amsmath}
\usepackage{amsfonts}
\usepackage{graphicx}
\usepackage{subfigure}
\usepackage{cite}
\usepackage{enumerate}
\usepackage{url}
\usepackage[ruled]{algorithm2e}
\usepackage{algorithmicx}
\usepackage{mathtools}
\usepackage{makecell}
\usepackage{slashbox} 
\usepackage{color, soul}
\allowdisplaybreaks[4]

\begin{document}

\newtheorem{definition}{\bf Definition}	
	
\title{\huge{Cooperation Techniques for A Cellular Internet of Unmanned Aerial Vehicles}}
\author{
\IEEEauthorblockN{
\normalsize{Hongliang Zhang}\IEEEauthorrefmark{1},
\normalsize{Lingyang Song}\IEEEauthorrefmark{1},
\normalsize{Zhu Han}\IEEEauthorrefmark{2},
\normalsize{and H. Vincent Poor}\IEEEauthorrefmark{3}\\}
\IEEEauthorblockA{\small
\IEEEauthorrefmark{1}School of Electrical Engineering and Computer Science, Peking University, Beijing, China.\\
\IEEEauthorrefmark{2}Electrical and Computer Engineering Department, University of Houston, Houston, TX, USA.\\
\IEEEauthorrefmark{3}Department of Electrical Engineering, Princeton University, Princeton, NJ, USA.}

}

\maketitle
\begin{abstract}
Unmanned aerial vehicles~(UAVs) are powerful Internet-of-Things components to provide sensing and communications in the air due to their advantages in mobility and flexibility. As aerial users, UAVs are envisioned to support various sensing applications in the next generation cellular systems, which have been studied by the Third Generation Partnership Project (3GPP). However, the Quality-of-Services (QoS) of the cellular link between the UAV and the base station may not be guaranteed when UAVs are at the cell edge or experiencing deep fading. In this article, we first introduce the non-cooperative cellular Internet of UAVs. Then we propose a cooperative sense-and-send protocol, in which a UAV can upload sensory data with the help of a UAV relay, to provide a better communication QoS for the sensing tasks. Key techniques including trajectory design and radio resource management that support the cooperative cellular Internet of UAVs are presented in detail. Finally, the extended cooperative cellular Internet of UAVs is discussed for QoS improvement with some open issues, such as massive multiple-input multiple-output systems, millimeter-wave, and cognitive communications.
\end{abstract}

\newpage

\section{Introduction}%

\begin{figure}[!t]
	\centering
	\includegraphics[width=5.0in]{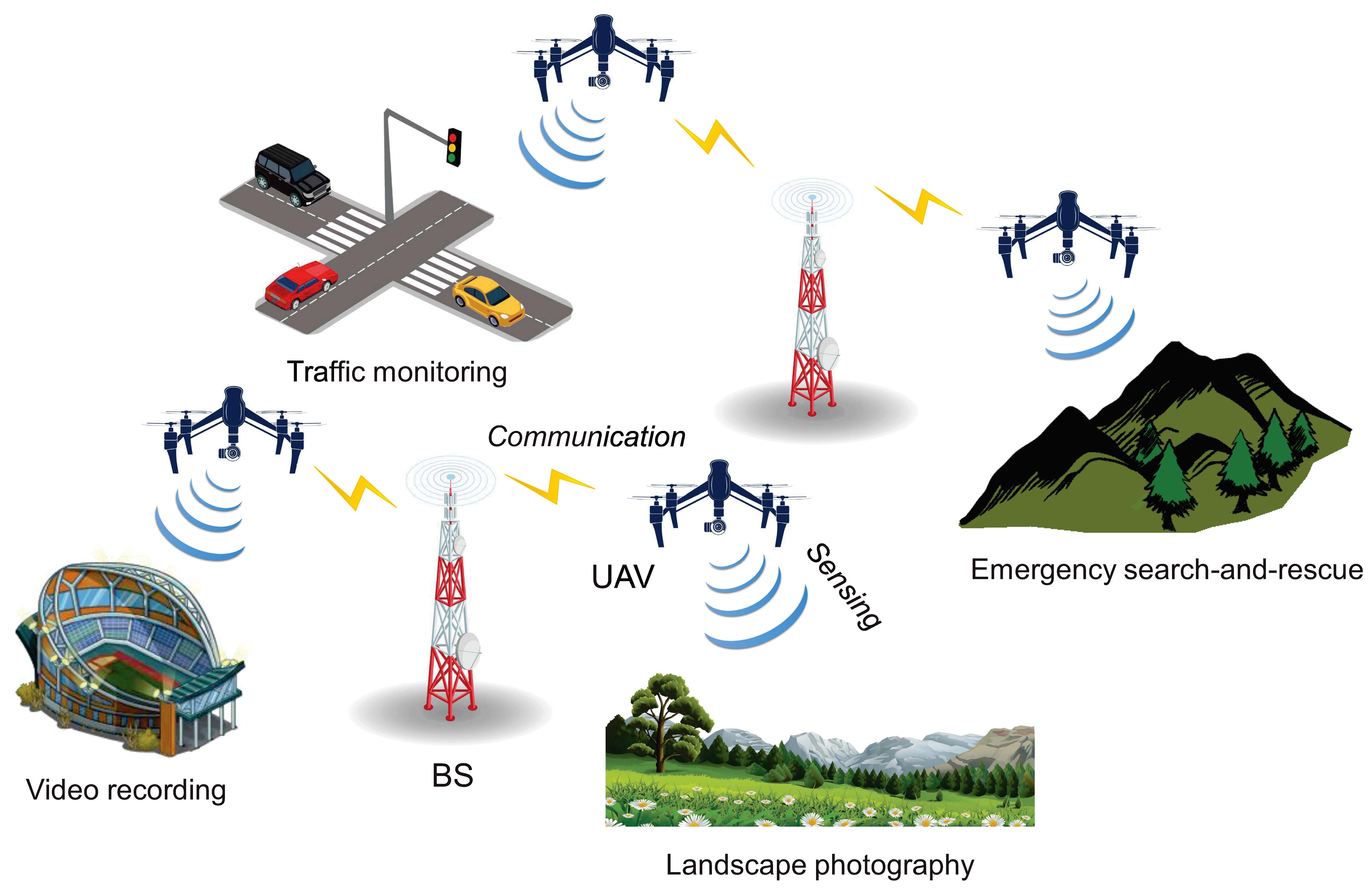}
	\caption{Illustration for UAV sensing applications.}
	\label{application}
\end{figure}

The easy availability of unmanned aerial vehicles~(UAVs) is generating considerable interest in civilian applications~\cite{SER-2016}, as illustrated in Fig.~\ref{application}. Recently, UAVs, equipped with cameras or sensors, have been widely used in our daily lives to execute a variety of sensing missions. Specially, UAVs are regarded as one of the best candidates to execute critical sensing missions such as emergency search-and-rescue and traffic monitoring due to their ease of deployment, high autonomy, and the ability to hover. Moreover, the low cost and high flexibility also make UAVs useful for the sensing applications with a huge volume of data, e.g., video recording and landscape photography. To turn the above technological visions into reality, seamless connectivity over wireless networking is necessary. Existing works focus on the ad hoc networking for UAV sensing~\cite{LRG-2016,ZAT-2018}. In~\cite{LRG-2016}, the authors surveyed the UAV ad hoc networks and discussed the routing and the handover among UAVs. In \cite{ZAT-2018}, an adaptive communication protocol was proposed to meet the demands of autonomy in airborne ad hoc networks for sensing applications. However, these ad hoc UAV networks are operated over unlicensed bands using CSMA/CA based protocols, which cannot guarantee the quality-of-service~(QoS) requirements. In addition, multi-hop routing among UAVs may lead to higher delay. Therefore, a more reliable network is necessary.

Recently, the Third Generation Partnership Project (3GPP) has approved the \emph{study item} on enhanced support to seamlessly integrate UAVs into future cellular networks~\cite{3GPP-2017}. Notably, it has produced accurate modeling of UAV-to-ground channels, and defined various UAV scenarios together with their respective features. Therefore, the terrestrial cellular networks, e.g., LTE and 5G, are considered as a promising enabler for UAV sensing applications, which we refer to as cellular Internet of UAVs. In the cellular Internet of UAVs, the sensory data needs to be transmitted to the base station~(BS) for timely processing. However, when UAVs are located at the cell edge or experiencing deep fading, the quality of the communication between the UAV and the BS may not be satisfactory, specially for emergency search-and-rescue and surveillance applications where the sensing task is far away from the BS. In the non-cooperative cellular Internet of UAVs, the UAV should fly to a communication point first to satisfy the QoS for the communication link, which will bring extra delay.

\begin{figure}[!t]
	\centering
	\includegraphics[width=5.0in]{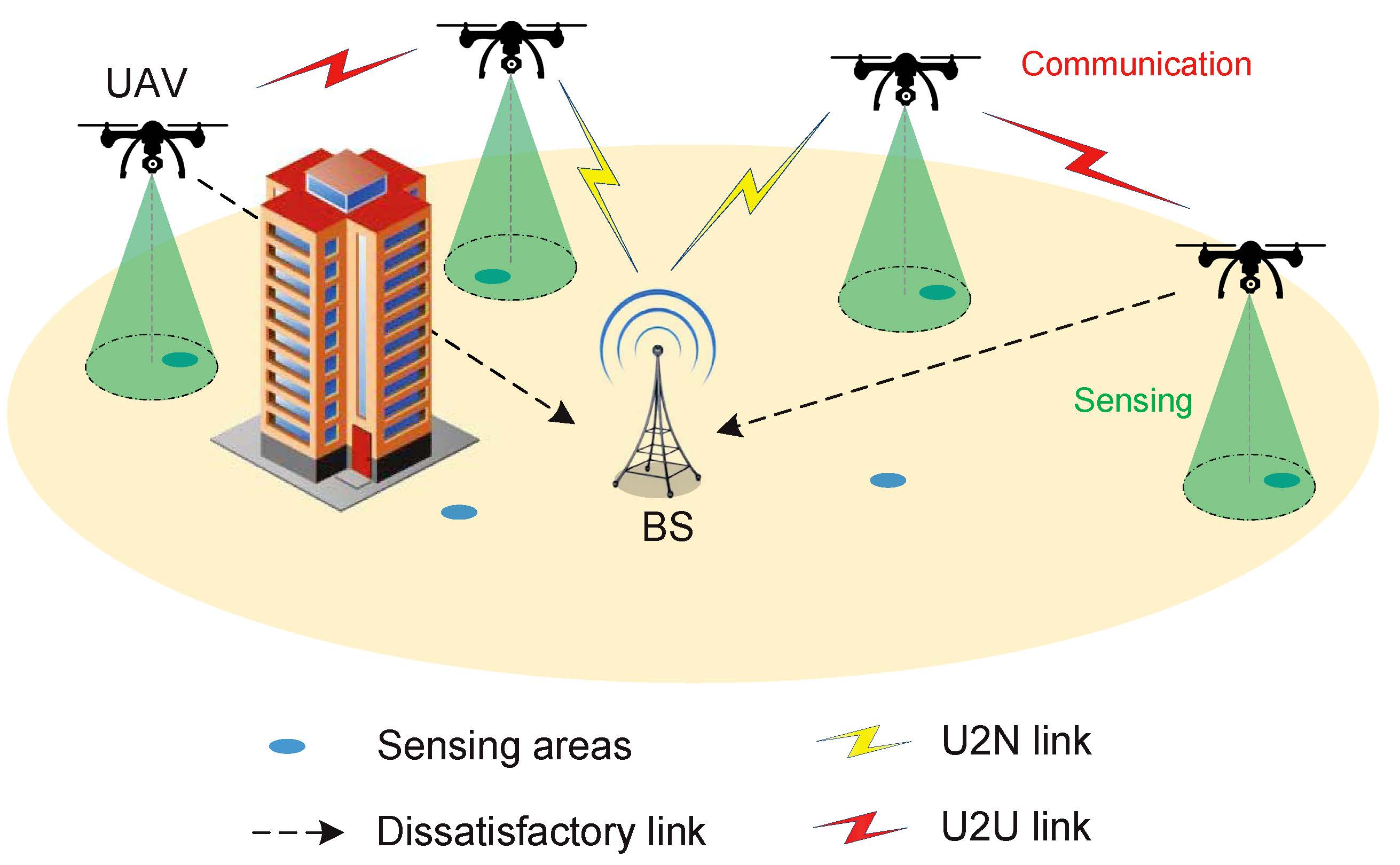}
	\caption{Illustration for cooperative cellular Internet of UAVs.}
	\label{Case}
\end{figure}

To tackle this challenge, we propose the UAVs to transmit via a UAV relay for providing a better QoS, as shown in Fig.~\ref{Case}. To support the cooperative cellular Internet of UAVs, the system provides two types of communications, i.e., UAV-to-network~(U2N) and UAV-to-UAV~(U2U). For U2N communications, the sensory data is sent to its destination by fiber in the backbone network instead of via the inefficient routing in the UAV ad hoc network, and thereby the latency becomes controllable. For U2U communications, two UAVs in proximity can set up a direct link bypassing the BS, data rates can be further improved by exploiting proximity gain and underlay nature. Unlike most previous works, which treat the UAVs as relays~\cite{SHQKL-2018} or flying BSs \cite{QYR-2018} and only consider the QoS for communications, we consider the UAVs to be aerial Internet-of-Things devices which keep executing sensing tasks when they work as mobile relays. Therefore, sensing and communications should be considered jointly.

In this article, we first illustrate a cooperative sense-and-send protocol. Then, based on the sense-and-send protocol, we present the key cooperation techniques for the cellular Internet of UAVs, which consist of trajectory design and radio resource management as follows.  

\begin{itemize}
	\item \textbf{Cooperative sense-and-send protocol:} The cellular UAVs need to transmit the sensory data to the BS immediately after each sensing task, and thus the sensing and communications are coupled. To facilitate the UAV cooperation, the procedure of sensing and communications should be well designed.
	
	\item \textbf{Cooperative trajectory design:} The UAVs perform sensing and communication simultaneously. Since the trajectory of the receiving UAV of a U2U link will have an impact on the transmitting UAV, it is necessary to design the trajectories jointly for these cooperative UAVs.
	
	\item \textbf{Cooperative radio resource management:} To further improve the spectrum efficiency, the U2U links can share the spectrum with the U2N ones. Therefore, the radio resource management, such as subchannel allocation and power control, is important to tackle the co-channel interference due to cooperation.
\end{itemize}   

In addition, we also discuss some cellular communication techniques that are applicable to cooperative cellular Internet of UAVs for QoS improvement in the future cellular networks, as illustrated below:

\begin{itemize}
	\item \textbf{Massive MIMO for U2N Communications:} Multiple antennas at the UAV can help improve the data rate by exploiting spatial diversity. Three-dimensional~(3D) beamforming techniques can create directional beams to reduce the inter-cell interference, which is especially suitable for U2N communications.
	
	\item \textbf{Millimeter-Wave for U2U Communications:}  Millimeter-Wave~(mmWave) band can significantly reduce the propagation loss but be sensitive to the blockage. Therefore, it can be utilized for U2U communications.
	
	\item \textbf{Cognitive Radio for Cooperative UAV Communications:} With cognitive radio~(CR), cooperative UAV communications can be extended to the spectrum occupied by the terrestrial cellular network to achieve higher data rates. 
\end{itemize}
 
The rest of this article is organized as follows. First, we introduce the basics of UAV sensing and communications in Section~\ref{Sec:2}. Then, key techniques for the cooperative cellular Internet of UAVs are presented in Section~\ref{Sec:3}. The main challenges, possible solutions, and performance evaluations are discussed in detail. With the aim to meet the strict QoS requirements in the future cellular networks, the extensions of the basic cellular Internet of UAVs are also discussed in Section~\ref{Sec:4} focusing on the corresponding relevant research problems. Finally, we draw our conclusions in Section~\ref{Sec:5}.

\section{Basics of UAV Sensing and Communications}
\label{Sec:2}

In Subsection \ref{2-1}, we first give a brief overview of the non-cooperative cellular UAV sensing. To provide a better QoS, we then introduce a cooperative sense-and-send protocol for the cellular Internet of UAVs in Subsection \ref{2-2}, where each sensing task consists of two parts: UAV sensing and UAV communications. Finally, we illustrate the channel models for UAV communications in Subsection \ref{2-3}. 

\subsection{Non-cooperative Cellular Internet of UAVs}
\label{2-1}

In the non-cooperative cellular Internet of UAVs~\cite{SHBL-S}, each UAV is required to execute several sensing tasks\footnote{The environment sensing is enabled by the sensors mounted on the UAVs \cite{YWHX-2018}.} independently and upload the sensory data by U2N communications concurrently\footnote{The UAV will not process the sensory data due to the limited computation capability. The collected data will be transmitted to the BS for further processing. The decision-making of the UAVs which have the capability to process the sensory data can be studied in the future work \cite{LKM-2018}.}. To avoid severe interference among different UAVs, different UAVs will use orthogonal subchannels for data transmission. Aiming to provide a high QoS for the sensing applications, we wish to minimize the completion time for the tasks\footnote{The completion time for the tasks is defined as the maximum time to complete the assigned tasks among UAVs.} while guaranteeing the QoS constraint for each U2N link. Due to the mobility of UAVs and the interactions between sensing and communication, trajectory design for the UAVs is important.

Define the location where the UAV performs sensing as the sensing point. The trajectory design can be performed for tasks successively, i.e., given the sensing point for the current task, the BS will determine the sensing point in the feasible area for the next task and the trajectory between these two sensing points. In the trajectory design, the time costs brought by UAV communication and UAV movement should be considered jointly. To be specific, the UAV may make a detour to a communication point with a better channel condition first and then move to the next sensing point after transmission. This is a non-convex continuous problem to find the communication point with both sensing and communication constraints, whose locally optimal solution can be obtained using gradient methods.     

\subsection{Cooperative Sense-and-Send Protocol}
\label{2-2}
To realize the cooperative UAV sensing and communications over cellular networks, we introduce a cooperative sense-and-send protocol~\cite{SHBL-2019}, which contains two phases: UAV sensing and UAV communications. Since the sensing and the communications are coupled by the trajectory of the UAVs, it is necessary to consider them jointly.

\begin{figure}[!t]
	\centering
	\includegraphics[width=5.0in]{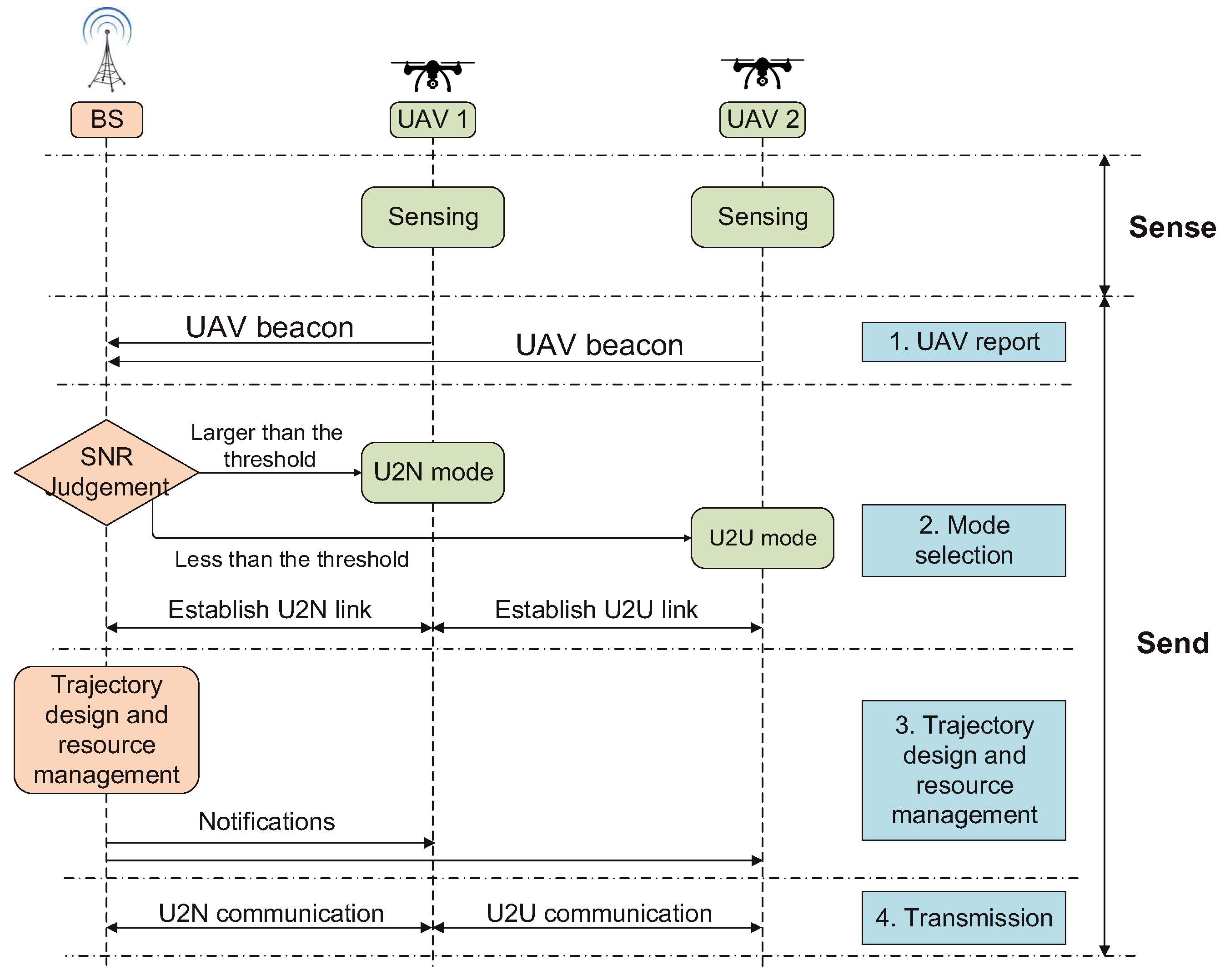}
	\caption{Cooperative sense-and-send protocol.}
	\label{protocol}
\end{figure}

\subsubsection{UAV Sensing} 

To guarantee the quality of sensing, the probability of UAV sensing failure needs to be less than a tolerance threshold. Define the feasible sensing area as the one where the requirement for sensing failure is satisfied. According to the probability model in \cite{VI-2017}, the probability of successful sensing is related to the distance between the location of the UAV and the center of the sensing area. Note that the UAV has a minimum flying height constraint, and thus the feasible sensing area for a task is a spherical crown. 

For a sensing task, the UAV hovers for a period over the sensing point and collects data. In the meanwhile, the UAV also transmits the sensory data to the BS. Note that a better sensing quality requires the UAV to get close to the sensing area while the UAV can obtain a better QoS for communications when it is closer to the BS. Therefore, the sensing and the communications have an impact on each other. 

\subsubsection{UAV Communications}

When performing data sensing, the UAV will upload the collected sensory data to the BS at the same time. However, the QoS requirements of the communication link may not be satisfied when the sensing task is located at the cell edge or the UAV suffers deep fading. To address this issue, the UAVs can transmit cooperatively, i.e., one UAV can transmit via another UAV as a relay. Therefore, the system provides two modes, i.e., U2N and U2U modes. 

\textbf{U2N mode:} U2N communications refer to the common cellular communications between the UAVs and the BS. Compared to UAV ad hoc networks, cellular transmissions can provide a more reliable data rate with a controllable radio environment. Moreover, instead of the multi-hop routing, the sensory data can be transmitted to the destination by fiber in the backbone network, and thus the latency becomes predictable. 

\textbf{U2U mode:} U2U communications refer to direct communications between two neighboring UAVs, bypassing the BS. The proximity afforded by U2U communications may provide a higher data rate between two UAVs, as the transmitted signal can suffer less path loss. Unlike traditional UAV relaying~\cite{SHQKL-2018}, the receiving UAV in the U2U communication also has its sensing tasks. 

As shown in Fig.~\ref{protocol}, the communication phase consists of four steps, i.e., UAV report, mode selection, trajectory design and resource management, and transmission. In the following, we will elaborate on these steps.

\textbf{UAV Report:} At the beginning of the communication phase, each UAV needs to send a beacon, which contains its ID and location, to the BS over the control channel in a time division manner.

\textbf{Mode Selection:} After receiving the beacon, the BS can determine the communication mode for each UAV according to the received signal-to-noise-ratio~(SNR). Specifically, if the received SNR is higher than the predefined threshold, the UAV transmits in the U2N mode, and otherwise, the UAV will transmit in the U2U mode. After receiving the notifications from the BS, the UAV in the U2N mode will establish a U2N link with the BS, and the UAV in the U2U mode will establish a U2U link with another UAV paired by the BS. 

\textbf{Trajectory Design and Resource Management:} The BS performs trajectory design and resource management, which will be presented in Section \ref{Sec:3}. Afterwards, the results are sent to the UAVs over the control channel.

\textbf{Transmission:} These UAVs start to transmit the sensory data to the corresponding BS or UAV over the allocated channel by the BS. The transmission step lasts until the end of this time slot.

\subsection{Channel Models}
\label{2-3}
Corresponding to the communication modes, there are two kinds of channels, i.e., U2N and U2U channels.

\textbf{U2N Channel:} The U2N channels significantly differ from terrestrial communication channels. Any movement caused by the UAVs can influence the channel characteristics. To be specific, the channel characteristics highly depends on the altitudes and elevation angles of the UAVs. A larger elevation angle will lead to a lower path loss with the same propagation distance. In addition, the channels are also affected by the propagation environment. Light-of-Sight~(LoS) links are expected in most scenarios, while they can also be occasionally blocked by obstacles such as terrain, buildings, or the airframe itself. 

To capture these features, a probability path loss model \cite{3GPP-2017} has been widely adopted for the U2N channel, in which the LoS and Non-Light-of-Sight~(NLoS) links are considered with different probabilities of occurrence. The probability model depends on the environment and elevation angle between the UAV and ground device. Moreover, the small-scale fading in the U2N channel can be characterized by the Rician fading model~\cite{DR-2017}, which consists of a deterministic LoS component, and a random scattered component.  

\textbf{U2U Channel:} The U2U channels are dominated by the LoS component. Although there may exist multipath fading due to ground reflections, its impact is limited compared to U2N channels as the altitude of the receiving UAVs is much higher than that of ground devices. In addition, the U2U channels may suffer more from Doppler frequency shifts than their U2N counterparts, due to the potentially large relative speed between two UAVs. Such channel characteristics can create significant challenges for spectrum allocation for U2U communications.

\section{Key Techniques for Cooperative Cellular Internet of UAVs}
\label{Sec:3}

Consider a single cell orthogonal frequency division multiplexing UAV network for cooperative cellular Internet of UAVs as shown in Fig.~\ref{Case}, where each UAV executes sensing and transmits the sensory data at the same time. Different from the non-cooperative cellular UAV sensing, the UAVs can either transmit the collected sensory data to the BS directly by U2N communications, or transmit the data through a UAV relay by U2U communications when the QoS for the direct communication link is not satisfactory~\cite{SHBL-2019}. Using the cooperative mode, the BS can extend its coverage and serve more UAVs. In the network, we assume that each U2U or U2N link is allocated to one subchannel for data transmission. In addition, to further improve the spectral efficiency, the U2U links work in an underlay mode, i.e., the U2U links can share the subchannel with U2N or other U2U links. 

To enhance the data traffic, we maximize the system sum-rate subjected to the QoS constraint for each communication link. Due to the cross-interference among U2N and U2U links, the network optimization becomes much more complicated than the non-cooperative scheme. In the following, we will elaborate on the problems of trajectory design in Subsection \ref{3-1} and radio resource management in Subsection \ref{3-2}, respectively.

\subsection{Cooperative Trajectory Design} 
\label{3-1}
The trajectory design consists of two parts: trajectory planning and speed control. The trajectory is planned for the tasks successively, in consideration of both sensing and communication performance. Given the sensing point for the current task, the sensing point for the next task is selected as the one in the feasible sensing area with the optimal U2N channel conditions, and the trajectory should be a line segment between these two sensing points.  

In addition, the speed of UAVs will influence the U2U communications, and thus speed control is necessary. If we do not consider the task completion time, the UAV trends to hover over the position with the optimal transmission condition. Therefore, we have a task completion time constraint in order to ensure that the UAV will fly forward towards the planned trajectory and complete the sensing tasks within the given time. Besides, each U2U link should satisfy a minimum transmission rate constraint. To tackle this problem efficiently, we control the speed in different time slots successively. Assume that the distance between a transmitting UAV and a receiving UAV is much larger than the maximum speed of a UAV, and thus the interference from other UAVs can be regarded as a constant. As the result, the speed control problem can be transformed into a convex problem and can be solved by existing convex techniques. Note that the speed of UAVs will be lower with a higher data rate for the U2U link. Therefore, the radio resource management also has an impact on the speed control.    

\subsection{Cooperative Radio Resource Management} 
\label{3-2}
The radio resource management includes subchannel allocation and power control: 

\textbf{Subchannel allocation:} The BS allocates the bandwidth resources to enhance the system sum-rate while guaranteeing the QoS constraint. Since a U2N link can share a subchannel with multiple U2U links, the co-channel interference needs to be considered here. Different from the conventional terrestrial cellular networks, the probabilistic radio propagation characteristics may pose important challenges on the spectrum management. Besides, the channel gain for a U2U link is also related to the trajectories of these two UAVs. Note that this is an NP-hard problem due to the binary variables. Therefore, to tackle this problem efficiently, we use the branch-and-bound method. The solution space for the problem can be regarded as a binary tree, where each node contains all the information of the variables. At the root node, all the variables are not fixed. An unfixed variable will be fixed to either 0 or 1 when a father node branches to two child nodes. The key idea of the branch-and-bound method is to prune the infeasible branches and approach the optimal solution efficiently. Initially, each UAV selects a set of subchannels based on its priority to satisfy the QoS constraint. In the following iterations, we first evaluate the bounds of the objective function and constraints separately to prune the branches that cannot achieve a better solution than the existing subchannel allocation result. Then, we fix an variable that has only one feasible value that satisfies the bound requirements. The iteration terminates until all the variables are fixed.

\textbf{Power control:} The transmit power for each UAV is controlled to further improve the system sum-rate, which is different from the non-cooperative scenario. Due to the existence of interference, the problem is a non-convex optimization problem. However, the data rate can be expressed as the difference of two convex functions, which can be solved by the difference of convex functions~(DC) algorithm. The basic idea is to successively approximate the feasible set by a sequence of polyhedral convex sets containing it and turn the problem into a series of convex problems. As such, we can obtain the local optimum efficiently. 

In summary, our proposed cooperative scheme is described below. In each time slot, each UAV sets up U2N or U2U communication link according to the cooperative sense-and-send protocol. Then, since trajectory design and radio resource management have impacts on each other, the BS performs the aforementioned trajectory design and radio resource management iteratively to maximize the system sum-rate, and the iterations stops when the objective function converges. 

\begin{figure}[!t]\label{cooperative} 
	\subfigure[]{
		\centering 
		\includegraphics[width=3.00in]{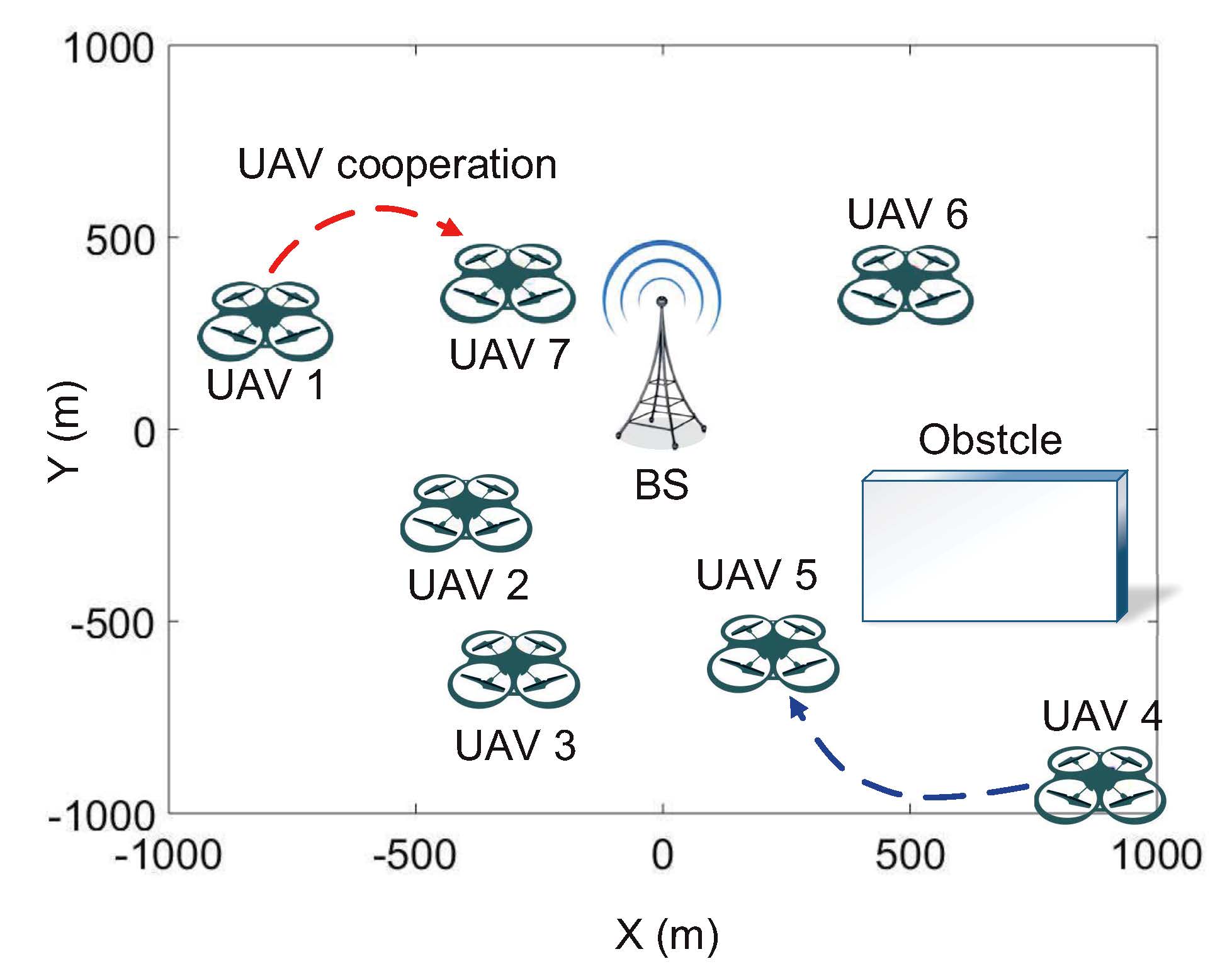} 
		\label{a}
	} 
	\subfigure[]{ 
		\centering 
		\includegraphics[width=3.4in]{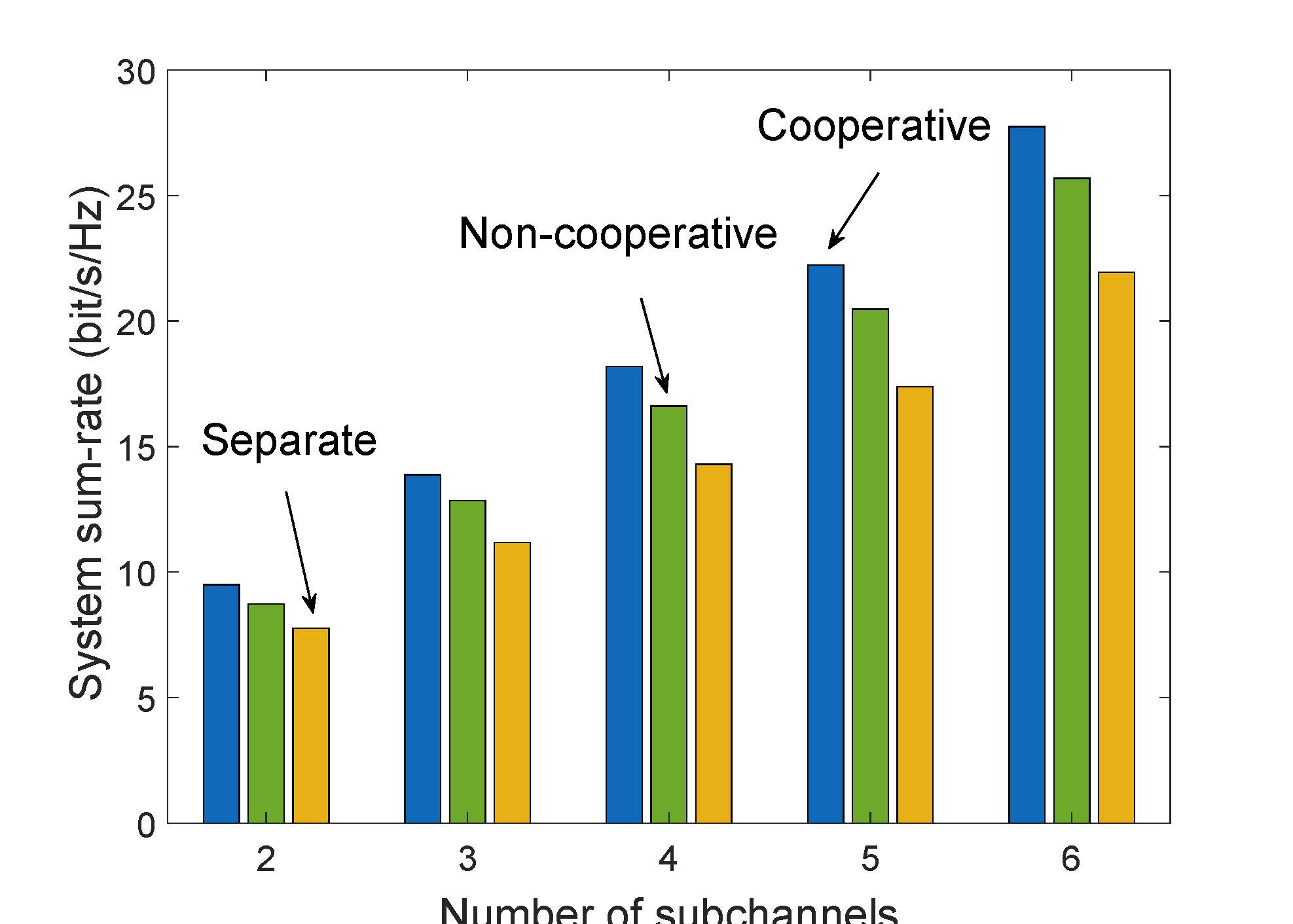} 
		\label{b}
	}
	\caption{Simulation evaluations: (a) A snapshot of a cooperative structure; (b) System sum-rate.} 
\end{figure} 

To evaluate the performance of the proposed cooperative scheme, we present a snapshot of the network obtained by the cooperative scheme in Fig.~\ref{a}, and compare it with the non-cooperative scheme and separate scheme\footnote{In the separate scheme, the UAVs will first fly to the best sensing location, and then fly to the nearest communication point where the QoS constraint can be satisfied.} in terms of the system sum-rate as shown in Fig.~\ref{b}. We assume that the SNR threshold for mode selection is $20$ dB and the UAVs select the nearest UAV that can satisfy the QoS requirements for U2N communications to serve as the UAV relay. We can observe that the cooperative scheme outperforms the non-cooperative one, especially when the subchannels is sufficient because the subchannels can be allocated to the U2U links more frequently. Besides, compared with the cooperative and non-cooperative schemes, the system sum-rate of the separate scheme is lower due to the extra flying time, which has shown the benefits brought by the joint optimization of sensing and communications.

\section{Cooperative Cellular Internet of UAVs Extensions for QoS Improvement}
\label{Sec:4}

To cope with the strict QoS requirements and expected immense amounts of traffic, future cellular networks are a mixture of different technologies. In this section, we will introduce some techniques applicable to the basic cooperative cellular UAV sensing for QoS improvement.

\subsection{Massive MIMO for U2N Communications}

\begin{figure}[!t]
	\centering
	\includegraphics[width=6.0in]{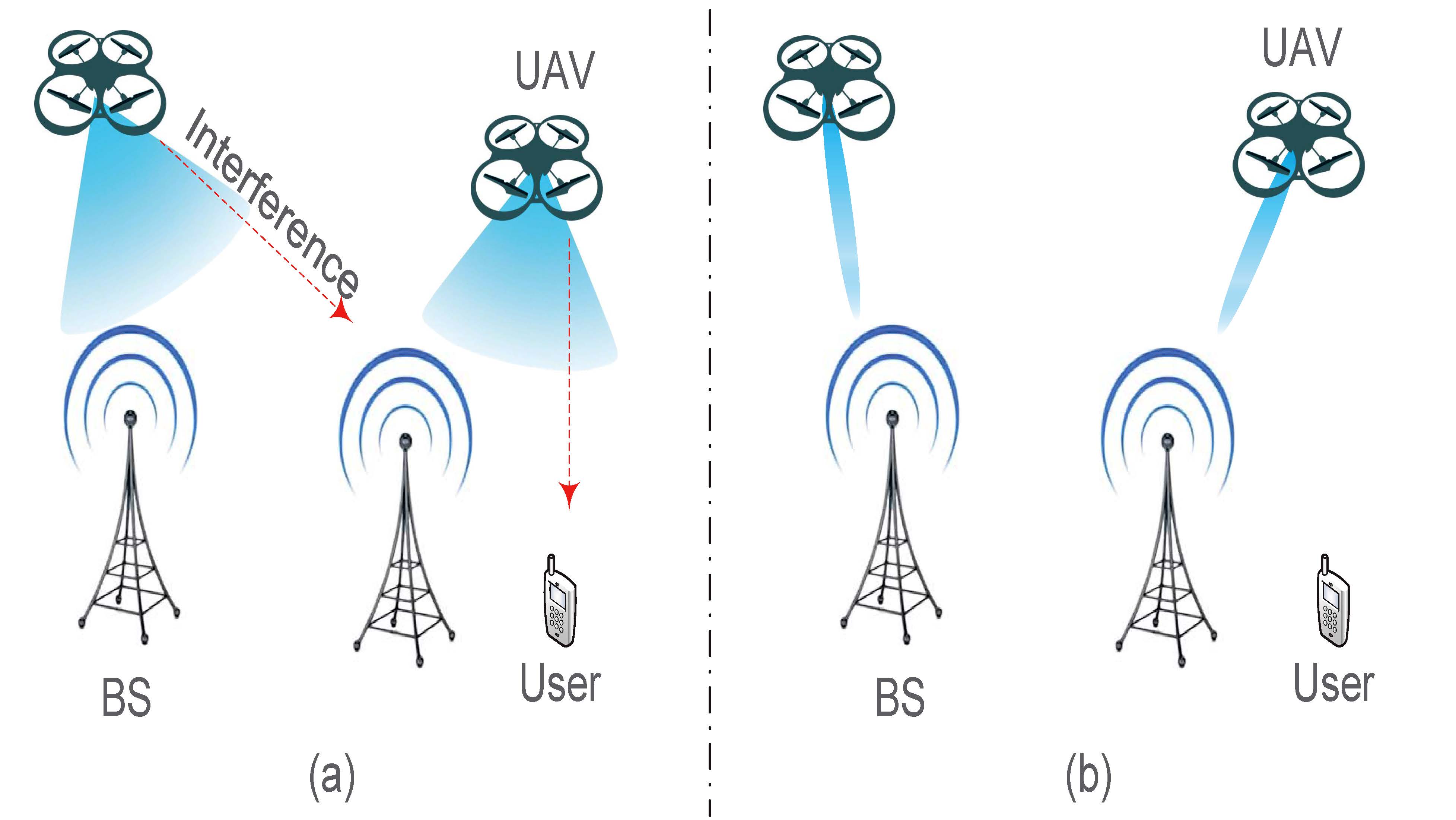}
	\caption{Integrations of massive MIMO into U2N communications: a) UAV equipped with a vertical antenna; b) UAV equipped with massive MIMO.}
	\label{MIMO}
\end{figure}

Massive multi-input multi-output (MIMO) is an emerging technique for 5G cellular wireless access, which is characterized by its scalability and potential to deliver very high and stable throughputs. With antenna arrays, 3D beamforming can exploit both the vertical and horizontal dimensions, and enable the creations of different beams in the space at the same time, thus reducing the interference~\cite{XBLJGYL-2014}. Due to the high altitude of the UAV and LoS dominant channel characteristic in U2N communications, ground users and neighboring BSs are vulnerable to be interfered with by a transmitting UAV equipped with a vertical antenna panel, as shown in Fig.~\ref{MIMO}(a). While employing massive MIMO and 3D beamforming, as shown in Fig.~\ref{MIMO}(b), the beam generated by a UAV equipped with a massive MIMO array can be BS-specific, thus mitigating the interference towards neighboring BSs and users. Therefore, massive MIMO and 3D beamforming facilitate the U2N communications.

The MIMO methods and possible research topics can be summarized as follows:

\begin{itemize}
	\item \textbf{UAV beamforming:} The UAV creates a 3D beam to avoid disturbing other U2N links. However, due to the mobility of the UAV, the precoding vectors need to be redeveloped.
	
	\item \textbf{Virtual UAV beamforming:} A swarm of UAVs can serve as a reconfigurable antenna array in the sky and collaboratively form beams to improve the system performance. Since the UAVs in the swarm may fly in different directions and speeds, the joint trajectory design will be very challenging in this situation.
	
	\item \textbf{BS multiplexing:} Multiple streams of UAVs coded with space-time codes, are transmitted through different antennas. Therefore, advanced detection schemes are necessary at the BS.
\end{itemize}

\subsection{Millimeter-Wave for U2U Communications}

The spectrum in the 3-300 GHz band, called mmWave, has a vast amount of unused or lightly used spectrum, which can be utilized to alleviate the spectrum shortage. However, the signals in this spectral band suffer a high path loss since the path loss grows with the square of the carrier frequency, and thus it is sensitive to blockage effects. Fortunately, the short wavelength of signals in this band allows a higher antenna gain with the same physical antenna size, and a directional antenna can be favored to compensate for the propagation loss and reduce shadowing effects. This indicates that mmWave communications are suitable to be employed in the scenarios without blockage, i.e., U2U communications.

The Doppler effect is one of the challenges brought by the mmWave implementation to the cooperative UAV communication systems. Since the Doppler effect depends on frequency and mobility, the Doppler shift will range from 10 Hz to 28 kHz if the mmWave frequency is 3-60 GHz with mobility speeds of UAVs within 3-500 km/h. Furthermore, due to the concentrated beam, there is a non-zero bias in the Doppler spectrum~\cite{LMFGP-2017}. Therefore, the Doppler effect of mmWave needs to be well addressed in both U2N and U2U communications. 

Another challenge is the handover strategy. When the UAV is flying away from the coverage of one mmWave radio and entering another one's, the UAV needs to hand over to avoid communication outage. However, there are two modes in the cooperative UAV communications, and thus, the handover complexity increases. In addition, because of the high mobility, directional transmission, and sensitivity to blockage, the UAV should also handover when the communication link is blocked by other UAVs.

\subsection{Cognitive Radio for Cooperative UAV Communications}

CR technology is an efficient approach to cope with the spectrum shortage and low utilization problems by opportunistic utilizing the frequency bands assigned to the primary users, subjected to the condition that transmissions of the primary users should not be interfered with by the secondary users~\cite{BK-2011}. For video recording and photographing applications, a large amount of sensory data collected by the UAVs imposes significant pressure on traditional cellular communications. To guarantee the demand of cellular communications while improving the quality of data transmission of UAVs, a CR access scheme can be used in which the UAVs can opportunistically access channels that are normally occupied by the cellular users. That is, cellular users and UAVs are regarded as primary users and secondary users, respectively. UAVs can only access the channel when the quality of cellular communications is not affected.

In this context, CR methods for enabling cooperative UAV communications can be studied in the following aspects. The first important issue is spectrum sensing. Through spectrum sensing, the UAVs can obtain information about its surrounding radio environment, which is the first critical step. However, the mobility of UAVs may result in false detection. Therefore, the UAVs can perform sensing cooperatively to reduce the false detection probability. The second one is spectrum handoff. Due to the dynamic topology, the spectrum handoff in CR UAV networks will be more frequent than other CR networks, which will degrade the system performance in latency and throughput. For this reason, a smooth spectrum handoff mechanism is necessary.

\section{Conclusions}
\label{Sec:5}

In this article, we have proposed cooperative cellular Internet of UAVs to provide guaranteed QoS for each sensing task. To facilitate the cooperation among these UAVs, the cooperative sense-and-send protocol consisting of UAV sensing and communications, has been illustrated. We then have addressed the key problems in cooperative cellular Internet of UAVs including trajectory design and radio resource management. The simulation results have shown that the cooperative scheme can further improve the system sum-rate compared with the non-cooperative one. Some applicable communication techniques to the basic cooperative cellular Internet of UAVs are also discussed, such as massive MIMO, mmWave, and CR, in order to cope with the explosive data traffic in the future cellular networks.  



\end{document}